# The influence of microwave radiation on the state of chromatin in human cells

Y.G. Shckorbatov [1], V.N. Pasiuga [1], V.A. Grabina[1], N.N. Kolchigin [1], D.O. Batrakov [1], V.V. Kalashnikov [2], D.D. Ivanchenko [1], V.N. Bykov [1]

[1] *Kharkiv National University, pl. Svobody, 4, Kharkiv, 61077, Ukraine,*
[2] *Kharkiv National University of Economics, Lenin Avenue, 9a, Kharkiv, 61001, Ukraine*

Isolated human buccal epithelium cells were irradiated by microwaves of frequency f=35 GHz and surface power density E=30 µW/cm$^2$. The state of chromatin in human cells was determined by methods of light and electron microscopy. The state of cell membranes was evaluated by the method of vital indigo carmine staining.

The microwave-induced condensation of chromatin in human cells was revealed. Left side circularly polarized waves induced less effect than linearly polarized radiation. The linearly polarized electromagnetic waves induced cell membrane damage revealed by increase of cell stainability. The data obtained are discussed in connection with the mechanisms of biological effects of electromagnetic waves.

## 1. Introduction

Artificial electromagnetic fields are the important component of environment in contemporary word. The mechanisms of biological action of this factor are not clear, but some data indicate potential hazard of low-level electromagnetic irradiation.

A study of military personnel in Poland showed a significantly increased relative risk of several nervous system tumors, including brain cancer, in persons exposed to electromagnetic fields of microwave diapason [1]. Odds ratios for both glioma and meningioma were slightly increased for long duration occupational exposure to microwave radiation [2]. A non-significantly increased risk of brain cancer was observed among men who had ever held a job with an average magnetic field exposure >0.6 µT relative to those with exposures <0.3 µT. A more pronounced risk was observed among men diagnosed with *glioblastoma multiforme*. Moreover, a cumulative time weighted index score of magnetic field exposure was significantly related to *glioblastoma multiforme* ($P = 0.02$) [3].

Such investigations are numerous, but very often the epidemiological studies do not provide a clear answer to the question of whether electromagnetic field (EMF) exposure influences the development of cancer. This is due in part to two shortcomings shared by epidemiological studies: *1)* small sample number and *2)* no specific EMF exposure parameter is known with which correlations should be made [4].

Great practical interest to the problem induces investigations of possible mutagenic effect of electromagnetic field of low intensity in radiofrequency diapason. Some investigations revealed mutagenic effect of radiofrequency irradiation. For instance in the work [5] the increase of chromosome aberrations and micronuclei in human lymphocytes under microwave irradiation (frequency 7.7 GHz, power density 0.5, 10 and 30 mW/cm$^2$, cell samples exposed for 10, 30 and 60 min.). Under extended exposure conditions, radio friquency signals at an average specific absorption rate (SAR) of at least 5.0 W/kg are capable of inducing chromosomal damage in human lymphocytes indicated by micronuclei test. Exposure to each of the four signals (ranged from 837 MHz to 1909.8 MHz) for 24 h at an average SAR of 5.0 or 10.0 W/kg resulted in a significant and reproducible increase of micronucleated lymphocytes frequency [6]. The micronucleus occurrence in Chinese hamster ovary (CHO)-K1 cells exposed to microwaves of frequency 2.45-GHz at a SAR lower than 50 W/kg did not differ from the sham-exposed controls, while those at SAR of 100 and 200 W/kg were significantly higher when compared with the sham-exposed controls. An increase of SAR causes a rise in temperature and this may be connected to the increase of micronuclei formation generated by exposure to microwaves [7].

But in other investigations no mutagenic effect of microwaves was demonstrated. For instance irradiation by 2.45 GHz microwaves for 2 h with up to 100 W/kg SAR with continuous wave-form and an average SAR of 100 W/kg with pulse wave-form (a maximum SAR of 900 W/kg) do not induce chromosomal aberrations in m5S mouse cells [8]. Also there was no evidence for induction of chromosome aberrations and micronuclei in human blood lymphocytes exposed in vitro for 24 h to 847.74 MHz radiation [9]. Irradiation with 847.74 - 813.56 MHz radiation that is applied in mobile phones at SAR 2.4 - 26 mW/kg not induce alterations in level of DNA damage or induce apoptosis in Molt-4 T lymphoblastoid cells [10].

In our previous works we show that in answer to microwave irradiation in buccal cell nuclei were formed heterochromatin granules [11,12] and draw the conclusion about microwave-induced chromatin condensation in human cell nuclei. For quantitative determination of heterochromatin changes resulted exposure to stress factors we introduced the abbreviation HGQ - heterochromatin granule quantity [13]. Cell nucleus electrokinetic properties change under the influence of microwave irradiation [11,14,15]. We also demonstrated increase of cell membrane permeability induced by microwave irradiation [14,15].

Later the fact of microwave-induced chromatin condensation was demonstrated by group of I.Beliaev using the method of chromatin anomalous viscosity time dependencies (AVTD). 30-min exposure to microwaves at 900 and 905 MHz resulted in statistically significant condensation of chromatin in human lymphocytes [16]. In the work [11] we demonstrated the HGQ increase after microwave irradiation of different circular polarization (f=42,25 GGz) and also by irradiation produced by cell phone. Group of I.Beliaev demonstrated the microwave-induced formation of foci containing tumor suppressor p53-binding protein 1 (53BP1) and phosphorylated histone H2AX (γ-H2AX) [17].

The purpose of the present work is to study the effects of low-level microwave radiation on human cells in connection with the state of circular polarization.

## 2. Materials and methods

### 2.1. Human Cells

Studies were realized in human cells of buccal epithelium. Cells were obtained from the inner surface of donor's cheek by light scraping with a blunt sterile spatula. This operation is absolutely bloodless and painless. All the good-will donors were informed about the purposes of investigation. Our investigations are performed in accordance with the European Convention on Human Rights and Biomedicine (1997), Declarations and Recommendations of the First, the Second and the Third National (Ukrainian) Congresses of Bioethics (Kiev, Ukraine, 2001, 2004, 2007) and Ukrainian legislation.

The cells were placed in solution of the following composition: 3.03 mM phosphate buffer (pH=7.0) with addition of 2.89 mM calcium chloride (*Reachem, Moscow, Russia*) and use for further experiments. 25 μl of cell suspension containing several thousand of cells were placed on the glass slide and subjected to microwave irradiation. Immediately after the irradiation procedure cells were stained with orcein or indigo carmine. Donors of cells were of male sex, non-smokers. Donor A was of 21 years old, donors B and C - 19 years, donor D – 35 years, and donor E – 51 years old.

### 2.2. Irradiation procedure

As a source of electromagnetic radiation of frequency f=35 GHz we applied a semi-conductor device. Irradiation was realized in a free space (10 cm from antenna edge). Irradiation was conducted at room temperature and no changes of the sample temperature during irradiation were registered.

In all experiments irradiation power density at the surface of exposed object was E=30 μW/cm$^2$. We applied linearly polarized and circularly polarized radiation. Irradiation time in all experiments was 10 s. The SAR of the cell suspension approximately equaled to that of water.

### 2.3. Chromatin state evaluation

In human cells we estimated the number of heterochromatin granules by the method described earlier [13]. The suspended cells (2 μl) were placed on the cover slide and irradiated. After irradiation cells were stained with 2 % orcein (*Merck AG, Darmstadt, Germany*) solution in 45 % acetic acid (*Reachem, Moscow, Russia*). Orcein is specific stain for heterochromatin staining as it was shown in a classic work [18]. Cells were investigated at magnification x 600. Each cell sample contained several thousands of cells. In each variant of experiment heterochromatin granules quantity (HGQ) was estimated in 30 cell nuclei and the mean HGQ value and the standard error of this value were calculated (presented in Fig.s 4-8). This number of cells (30) was determined in our previous experiments as an optimal for such analysis. The variability of (HGQ) in cell population gives the value of the standard error of the mean data (SEM) less than 5% of the mean HGQ that is enough for biological experiments. Three independent experiments for each donor was conducted using three different cell samples obtained in different days.

### 2.4. Evaluation of the state of cell membrane

We applied indigo carmine as cell damage indicator. This method also may be considered as a method reflecting cell viability. Previously it was shown that cell damage induces the increase of percentage of stained cell [19]. Therefore we used the percentage of unstained cells after 5 min of staining with 5 mM indigo carmine solution in the buffer solution described above. In one experiment we analyzed 3 cell samples irradiated independently ($N_1$). In each cell sample we analyzed 100 cells ($N_2$) and determined the percentage of unstained cells for each cell sample. After this we calculated the mean number of unstained cells for 3 experiments and the standard error of this value (presented in the Fig. 9).

## 2.5. Image processing

For more distinct determination of heterochromatin location in interphase cell nucleus we elaborated computer program that enabled to process the digital images and to paint zones with different heterochromatin contents with different colors. In the Fig.1 are presented the results of image processing in colorless variant.

The elaborated methods of computer visualization make the work of microscopic analysis much easier and give the opportunity to avoid the mistakes resultant from personal subjectivity.

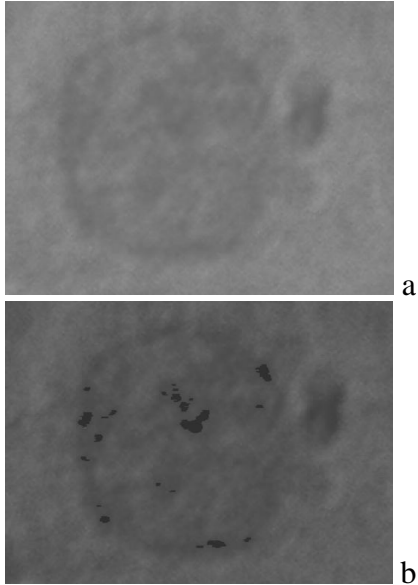

**Fig. 1. The image of human buccal epithelium cell nucleus processed by an original computer program: before (a) / after (b) the image processing (magnification x 600).**

## 2.6. Electron microscopy

The electron microscopy investigations were realized on microscope EM-125 ("Electron" factory, city of Sumiy, Ukraine).

## 2.7. Statistical analysis

All experimental results were statistically processed using Student's t-test. The probability level assumed in this paper is $P<0,05$. Significantly reliable changes of control are marked in Fig. 4 – 9 with asterisks (*).

# 3. Results

Electron microscopic image of the nucleus of buccal epithelium is presented in Fig. 2. After microwave irradiation the chromatin condensation located by the main part near nuclear envelope is observed (Fig. 3).

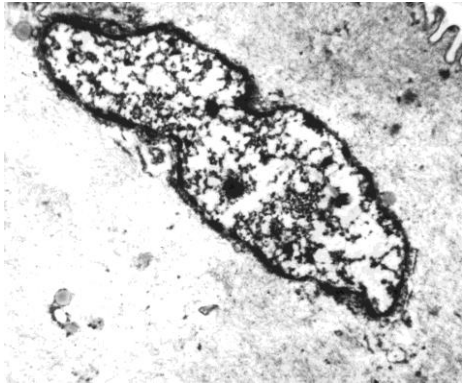

**Fig. 2. The nucleus of buccal epithelium cell at magnification x 12000.**

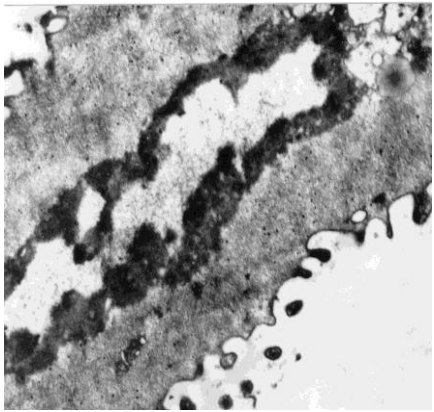

**Fig. 3. The nucleus of buccal epithelium cell after microwave irradiation (power density 200 μW/cm$^2$, irradiation time 60 s, magnification x 12000).**

The microwave irradiation of human cells induces the significant increase of HGQ parameter. As one can see in Figs 4-8 cell exposure during 10 s induce increase of HGQ. This increase was registered in all tested donors in no relation to initial HGQ level. In cells of elder donor E the initial level of HGQ was higher than in cells of other donors that is in a good agreement with our previous results indicating age related condensation of chromatin [20].

As one can see from the presented data, almost in all experiments the right and left polarized microwaves induced approximately equal biological effect, but the left polarized electromagnetic waves induced less biological effect than linearly polarized ones (P<0,05).

The applied intensity of irradiation induces cell damage that is manifested by decreasing of percentage of unstained cells after cell exposure to linearly polarized microwaves (Fig. 9).

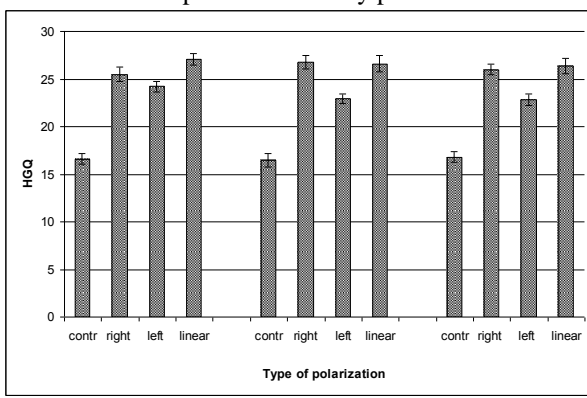

**Fig. 4. Changes in heterochromatin granule quantity (HGQ) after cell exposure to differently polarized microwaves in cells of donor A. Error bars indicate the standard error of the mean (SEM) for N = 30 independent experiments.**

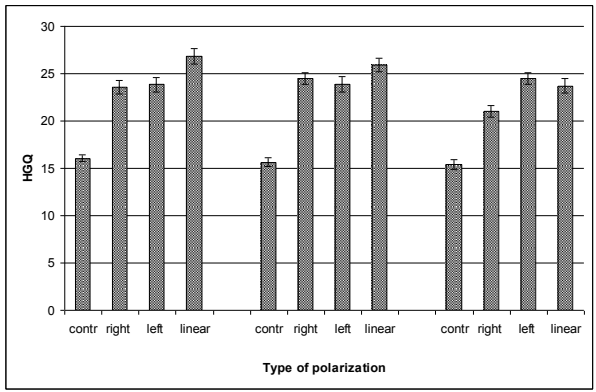

**Fig. 5.** Changes in heterochromatin granule quantity (HGQ) after cell exposure to differently polarized microwaves in cells of donor B. Error bars indicate the standard error of the mean (SEM) for N = 30 independent experiments.

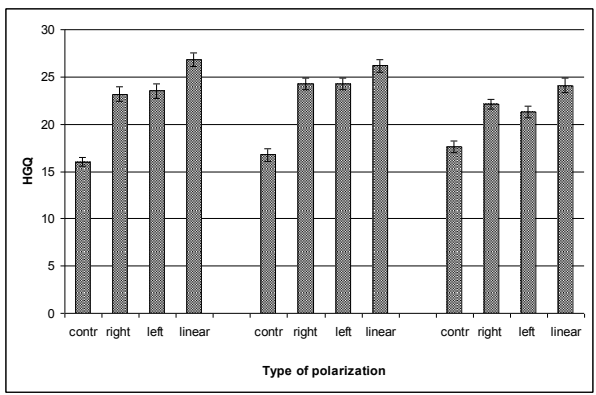

**Fig. 6.** Changes in heterochromatin granule quantity (HGQ) after cell exposure to differently polarized microwaves in cells of donor C. Error bars indicate the standard error of the mean (SEM) for N = 30 independent experiments.

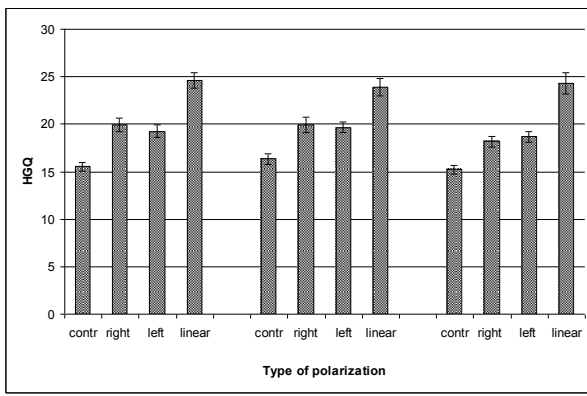

**Fig. 7.** Changes in heterochromatin granule quantity (HGQ) after cell exposure to differently polarized microwaves in cells of donor D. Error bars indicate the standard error of the mean (SEM) for N = 30 independent experiments.

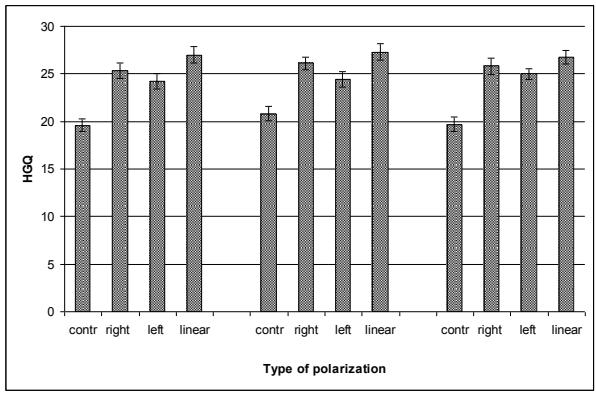

**Fig. 8. Changes in heterochromatin granule quantity (HGQ) after cell exposure to differently polarized microwaves in cells of donor E. Error bars indicate the standard error of the mean (SEM) for N = 30 independent experiments.**

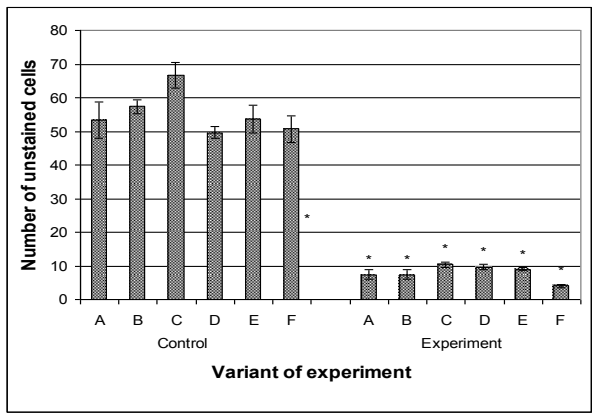

**Fig. 9. Changes in cell membrane permeability after cell exposure to linearly polarized microwaves. Error bars indicate the standard error of the mean (SEM) for $N_1$ = 3 independent experiments. The number of analyzed cell in each experiment was - $N_2$=100.**

## 4. Discussion

The mechanism of biological action of microwaves is not yet known. Different experimental data indicate the important role of DNA and genome in it. Electromagnetic radiation induces heat shock factor activation [21] connected with EMF induction of field-responsive domain in the heat-shock protein 70 (HSP-70) promoter [22]. These and others experimental facts induced the hypothesis about general mechanism of EMF action via electromagnetic initiation of transcription at specific DNA sites [23]. This hypothesis is based on the notions concerning the interaction of external electromagnetic fields immediately with DNA electrons [24]. In our previous work we demonstrated that microwave radiation induces chromatin condensation adjusting nuclear envelope [11]. Our present data indicate chromatin condensation in the whole nucleus. We suppose that chromatin condensation is general answer to EMF irradiation. Chromatin condensation maybe induced by changing of DNA protein interaction evoked by electromagnetic filed. This possibility is proofed in the work [25] in which changing of protein DNA interaction was shone for regulatory proteins of chromatin. Possibly chromatin condensation may be a cause of mutations because it is known that heterochromatic state (chromatin condensation) in chromosomes leads to mutability increase [26].

We have demonstrated increase of number of heterochromatin granules in response to microwave irradiation. The formation of such granules in response to action of stress factors was previously demonstrated in other experimental systems. The concentration of heat shock factor 1 (HSF1) in the stress-induced interphase chromatin granules (HSF1 granules) was shown in the work [27]. HSF1 stress granules were detected within 30 seconds of heat shock. HSF1 stress granules were detected after 5-10 minutes of treatment with cadmium, and steady state was reached after about 24 minutes of cadmium exposure [26]. We suppose that microwaves induce cell response to stress which is manifested in the process of heterochromatinization. It is known that the process of heterochromatinization is connected with decrease of transcriptional activity [28]. In our experiments microwaves induced cell membrane damage revealed by decrease of unstained by indigo carmine cells. This phenomenon supports the view that low-level microwave irradiation induces stress reaction in human cells.

Our experimental data on different sensibility of cells to differently polarized microwave irradiation may be interpreted in connection with asymmetry of biological molecules, first of all DNA. It is known that DNA molecule is a right helix and therefore its interaction with differently circularly polarized microwaves may be the result of DNA asymmetry. On the further stages of the reaction to electromagnetic irradiation it may result in differences of the process of microwave-induced heterochromatinization.

## 5. Conclusions

The data obtained in this work demonstrate important biological effects of monochromatic microwave irradiation. Low-level microwave irradiation induces chromatin condensation in human cells and damage of cell membranes. The left circularly polarized microwaves induced less chromatin condensation than linearly polarized ones.